# The Effect of Stay-at-Home Orders on COVID-19 Cases and Fatalities in the United States

James H. Fowler[1,2,*], Seth J. Hill[2], Remy Levin[3], Nick Obradovich[4]

[1] Infectious Diseases and Global Public Health Division, University of California, San Diego
[2] Political Science Department, University of California, San Diego
[3] Economics Department, University of California, San Diego
[4] Center for Humans and Machines, Max Planck Institute for Human Development
* corresponding author, fowler@ucsd.edu

## Abstract

Governments issue "stay at home" orders to reduce the spread of contagious diseases, but the magnitude of such orders' effectiveness is uncertain. In the United States these orders were not coordinated at the national level during the coronavirus disease 2019 (COVID-19) pandemic, which creates an opportunity to use spatial and temporal variation to measure the policies' effect with greater accuracy. Here, we combine data on the timing of stay-at-home orders with daily confirmed COVID-19 cases and fatalities at the county level in the United States. We estimate the effect of stay-at-home orders using a difference-in- differences design that accounts for unmeasured local variation in factors like health systems and demographics and for unmeasured temporal variation in factors like national mitigation actions and access to tests. Compared to counties that did not implement stay-at-home orders, the results show that the orders are associated with a 30.2 percent (11.0 to 45.2) reduction in weekly cases after one week, a 40.0 percent (23.4 to 53.0) reduction after two weeks, and a 48.6 percent (31.1 to 61.7) reduction after three weeks. Stay-at-home orders are also associated with a 59.8 percent (18.3 to 80.2) reduction in weekly fatalities after three weeks. These results suggest that stay-at-home orders reduced confirmed cases by 390,000 (170,000 to 680,000) and fatalities by 41,000 (27,000 to 59,000) within the first three weeks in localities where they were implemented.

# Introduction

Coronavirus disease 2019 (COVID-19) first appeared as a cluster of pneumonia cases in Wuhan, China on December 31, 2019[1] and was declared a global pandemic by the World Health Organization (WHO) on March 11, 2020.[2] As of May 6, 2020, the European Centers for Disease Control reports that worldwide there have been 3,623,803 confirmed cases of COVID-19, resulting in 256,880 deaths.[3]

The United States has both the highest number of cases (1,204,475) and deaths (71,078) due to the disease.[3] As a result, the U.S. government has been widely criticized for inaction in the early stages of the pandemic.[2] Although the first confirmed case of COVID-19 was reported to the Centers for Disease Control on January 21, 2020 and documented transmission commenced immediately,[4] a national state of emergency was not declared until nearly two months later on March 13. At that time, the only mandatory action at the national level was international travel restrictions.[5]

While the national government has the authority to act, the United States is a federal political system where public health is normally the purview of the fifty states. Furthermore, each state often delegates health authority to cities and/or counties, geographic political units nested within states. As a result, responses to COVID-19 varied across states and counties and led to spatial and temporal variation in implementation of mitigation procedures. This variation in policy responses has likely contributed to significant variation in the incidence of cases and fatalities -- as well as the downstream social and economic effects of the disease -- across jurisdictions in the United States.[6–11]

A variety of government policies have been proposed and used to mitigate the spread and consequence of pandemic diseases like COVID-19, ranging from investments in medical testing, contact tracing, and clinical management, to school closures, banning of mass gatherings, quarantines, and population stay-at-home orders.[12] China's extensive interventions appear to have been successful at limiting the outbreak.[13,14] These include quarantines both for those diagnosed and those undiagnosed but who had been in Hubei province during the outbreak,[15] and restrictions on travel to and from affected areas.[16] In contrast, school closures across East Asia were estimated to be much less effective.[17]

With estimates that nearly half of transmissions occur from pre-symptomatic and asymptomatic individuals, epidemiological simulations suggest that quarantines of symptomatic individuals alone will be insufficient to halt the pandemic.[18] This has led to widespread adoption of population-wide policies to dramatically reduce social contact.

Here, we study the role of stay-at-home orders, perhaps the most common policy intervention in the United States and Europe. Stay-at-home orders require citizens to shelter in their residence with very few exceptions and have typically been implemented along with school closures, bans on mass gatherings, and closure of non-essential businesses. These policies are associated with a significant reduction in observed mobility,[19] and initial evidence from New York City and California suggests that they can be effective in reducing case growth in the United States.[8,20] Yet, because each locality in the U.S. has many factors that contribute to differential rates of transmission, statistical efforts to control for potential confounds and to identify the precise effects of stay-at-home orders across the U.S. are critical to understanding whether -- and to what degree -- such policies are working.

## Results

We collected data on stay-at-home orders, COVID-19 confirmed cases, rates of testing, and fatalities by day and county throughout the United States (see Methods). Figure 1 shows that the number of cases and fatalities grew exponentially from March 1 to May 3, 2020. It also suggests that efforts to "flatten the curve" initiated in mid-to-late March helped to reduce the rate of exponential growth.

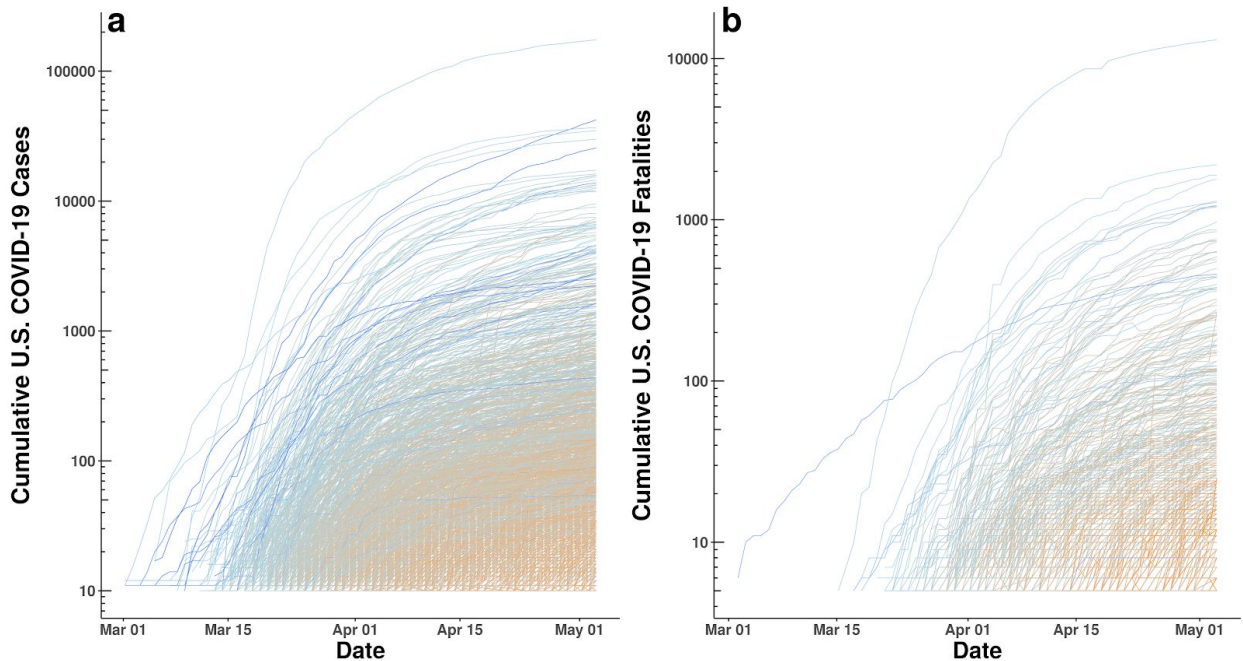

**Figure 1. (a)** Log of cumulative confirmed COVID-19 cases by county and date. **(b)** Log of cumulative confirmed COVID-19 fatalities by county and date. Lines are gradient-colored by date of first case and date of first fatality, respectively (blue = early to light blue = middle to orange = late).

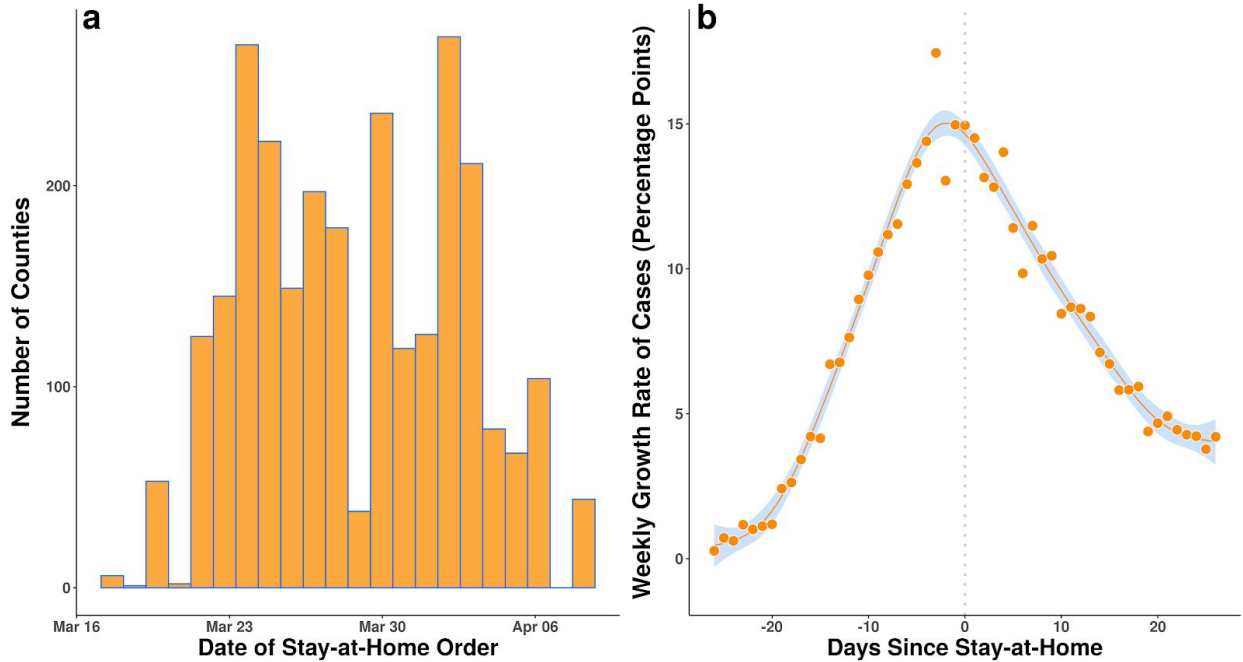

**Figure 2. (a)** Distribution of stay-at-home orders at the county level by date in the United States. **(b)** Mean county-level weekly growth in total confirmed COVID-19 cases in the United States by the number of days before or after a stay-at-home order.

Figure 2a shows the distribution of stay-at-home order dates. By April 7, 2020, 18 states (1,451 counties) had different counties with different order dates, 27 states and the District of Columbia had statewide orders with no local variation (1,307 counties) and 5 states (386 counties) had no order in place. Figure 2b shows how the mean county-level weekly growth rate in COVID-19 cases changed relative to the date the stay-at-home order went into effect in counties that implemented a stay-at-home order. Peak growth in these counties occurred three days before the order went into effect.

The growth rate of cases began to decline following the orders. In Figure 3, we group counties by the date a stay-at-home order was implemented and show how confirmed cases and fatalities changed with respect to the order. For example, the top blue curve in both panels refers to all counties with stay-at-home orders enacted on March 23 or 24, notably including the epicenter of the pandemic in New York. We then plot the set of counties that did not issue any stay-at-home order during this time period with a thicker orange line to highlight how this group of counties differs from those with an order. Although the curves have been "flattened" in all counties, the flattening in counties without an order is much less, particularly with respect to cases.

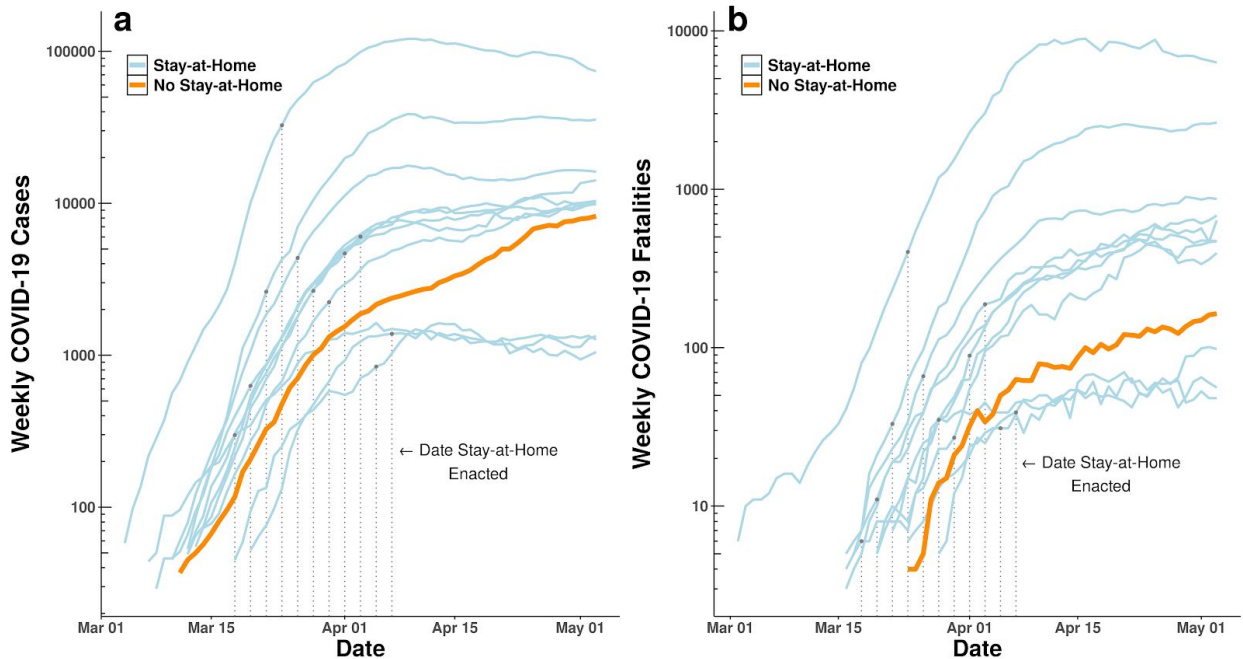

**Figure 3.** Weekly **(a)** newly-confirmed COVID-19 cases in the United States and **(b)** confirmed fatalities. Each line represents an aggregation of counties that issued stay-at-home orders during the same two-day period (blue) or did not issue an order (orange). Observations are grouped to two-day pairs to facilitate visualization.

Table 1 shows growth in log weekly cases both on the date an order goes into effect and 21 days after the order. Each row presents corresponding weekly growth on the same dates for the set of all counties where no such order ever went into effect. For example, the first row shows that weekly cases for the six counties who issued an order on March 17 grew from March 10 to March 17 by 1.02 units on a log scale. For the 506 counties that would never issue such an order, weekly case growth from March 10 to March 17 averaged 1.39 units on a log scale.

Although growth slowed for all counties over each 21 day period, with or without an order, the last column in Table 1 shows that it slowed faster for every single county group with an order than for the counties that did not issue an order. Fatalities follow the same pattern, as shown in Table 2. Once again, growth slows for all counties, but the rightmost column shows that 1,931 of the 2,647 counties with orders are in groups that decline faster than counties without orders.

Although these results suggest that stay-at-home orders worked, a number of factors might confound this association in the raw data. For example, stay-at-home orders might closely follow earlier targeted mitigation measures at the national level (such as travel restrictions issued by the State Department or recommendations by the CDC on mass gatherings). There might also exist

|  | | One Week Change in Log Weekly Cases | | | | | | |
|---|---|---|---|---|---|---|---|---|
|  | | Counties With Stay-at-Home Order | | Counties Without Stay-at-Home Order | | | | |
| Date Stay-at-Home Order Went Into Effect | Number of Counties | On Day of Order | 21 Days After Order | On Day of Order | 21 Days After Order | 21-Day Difference in Change, With Order | 21-Day Difference in Change, Without Order | Difference in Difference, Counties With Orders and Without Orders |
| 3/17/20 | 6 | 1.02 | 0.71 | 1.39 | 1.10 | -0.31 | -0.29 | -0.02 |
| 3/19/20 | 52 | 2.00 | 1.00 | 1.34 | 0.89 | -1.00 | -0.45 | -0.55 |
| 3/21/20 | 104 | 2.52 | 0.88 | 1.52 | 0.66 | -1.64 | -0.85 | -0.79 |
| 3/22/20 | 23 | 2.87 | 0.99 | 1.59 | 0.66 | -1.88 | -0.92 | -0.96 |
| 3/23/20 | 209 | 2.63 | 0.41 | 1.50 | 0.54 | -2.22 | -0.96 | -1.26 |
| 3/24/20 | 258 | 2.20 | 0.57 | 1.61 | 0.50 | -1.63 | -1.10 | -0.53 |
| 3/25/20 | 308 | 1.93 | 0.47 | 1.65 | 0.46 | -1.46 | -1.19 | -0.27 |
| 3/26/20 | 77 | 1.67 | 0.22 | 1.42 | 0.39 | -1.45 | -1.02 | -0.43 |
| 3/27/20 | 136 | 1.52 | 0.16 | 1.42 | 0.34 | -1.36 | -1.08 | -0.28 |
| 3/28/20 | 174 | 1.40 | 0.33 | 1.35 | 0.33 | -1.07 | -1.02 | -0.04 |
| 3/29/20 | 38 | 1.32 | -0.10 | 1.23 | 0.25 | -1.41 | -0.98 | -0.44 |
| 3/30/20 | 324 | 1.35 | 0.32 | 1.29 | 0.28 | -1.03 | -1.01 | -0.02 |
| 3/31/20 | 31 | 1.54 | 0.58 | 1.10 | 0.28 | -0.96 | -0.82 | -0.14 |
| 4/1/20 | 126 | 1.10 | -0.13 | 0.93 | 0.31 | -1.24 | -0.62 | -0.61 |
| 4/2/20 | 274 | 1.13 | 0.17 | 0.89 | 0.30 | -0.97 | -0.58 | -0.38 |
| 4/3/20 | 290 | 0.83 | 0.08 | 0.78 | 0.32 | -0.75 | -0.46 | -0.29 |
| 4/4/20 | 66 | 0.54 | -0.04 | 0.66 | 0.38 | -0.59 | -0.29 | -0.30 |
| 4/6/20 | 104 | 0.58 | -0.17 | 0.54 | 0.40 | -0.76 | -0.14 | -0.61 |
| 4/7/20 | 44 | 0.61 | -0.05 | 0.50 | 0.46 | -0.66 | -0.04 | -0.62 |

**Table 1.** Comparison of weekly changes in log daily confirmed cases of COVID-19 between 2,647 counties with stay-at-home orders and 506 counties without orders on the same date. Although mean rates of change decline for all counties, the rightmost column shows that all of the 2,647 counties with orders are in groups that decline faster than counties without orders.

|  | | One Week Change in Log Weekly Fatalities | | | | | | |
|---|---|---|---|---|---|---|---|---|
|  | | Counties With Stay-at-Home Order | | Counties Without Stay-at-Home Order | | | | |
| Date Stay-at-Home Order Went Into Effect | Number of Counties | On Day of Order | 21 Days After Order | On Day of Order | 21 Days After Order | 21-Day Difference in Change, With Order | 21-Day Difference in Change, Without Order | Difference in Difference, Counties With Orders and Without Orders |
|---|---|---|---|---|---|---|---|---|
| 3/17/20 | 6 | 1.10 | 0.12 | -0.69 | 0.94 | -0.98 | 1.63 | -2.62 |
| 3/19/20 | 52 | 1.20 | 0.76 | -0.69 | 0.43 | -0.45 | 1.12 | -1.57 |
| 3/21/20 | 104 | 2.56 | 0.82 | -0.69 | 0.71 | -1.74 | 1.40 | -3.14 |
| 3/22/20 | 23 | 1.85 | 0.63 | 0.00 | 0.40 | -1.21 | 0.40 | -1.61 |
| 3/23/20 | 209 | 1.81 | 0.33 | 0.00 | 0.34 | -1.48 | 0.34 | -1.82 |
| 3/24/20 | 258 | 3.62 | 0.61 | 1.61 | 0.16 | -3.01 | -1.45 | -1.56 |
| 3/25/20 | 308 | 2.01 | 0.43 | 1.61 | 0.33 | -1.59 | -1.28 | -0.31 |
| 3/26/20 | 77 | 1.61 | 0.10 | 1.79 | 0.47 | -1.51 | -1.32 | -0.19 |
| 3/27/20 | 136 | 1.10 | -0.14 | 2.48 | 0.16 | -1.24 | -2.32 | 1.08 |
| 3/28/20 | 174 | 2.25 | 0.28 | 2.71 | 0.29 | -1.98 | -2.41 | 0.44 |
| 3/29/20 | 38 | 1.50 | 0.12 | 2.08 | 0.26 | -1.39 | -1.82 | 0.43 |
| 3/30/20 | 324 | 0.80 | 0.34 | 2.40 | 0.31 | -0.46 | -2.09 | 1.63 |
| 3/31/20 | 31 | 1.28 | 0.01 | 1.61 | 0.49 | -1.27 | -1.12 | -0.14 |
| 4/1/20 | 126 | 1.72 | 0.05 | 1.89 | 0.32 | -1.67 | -1.57 | -0.10 |
| 4/2/20 | 274 | 1.66 | 0.44 | 1.92 | 0.16 | -1.22 | -1.76 | 0.54 |
| 4/3/20 | 290 | 1.12 | 0.06 | 1.07 | 0.34 | -1.06 | -0.73 | -0.33 |
| 4/4/20 | 66 | 1.86 | 0.27 | 0.96 | 0.18 | -1.59 | -0.77 | -0.81 |
| 4/6/20 | 104 | 0.77 | -0.13 | 0.92 | 0.23 | -0.90 | -0.69 | -0.21 |
| 4/7/20 | 44 | 0.65 | 0.40 | 0.94 | 0.02 | -0.26 | -0.92 | 0.67 |

**Table 2.** Comparison of weekly changes in log daily fatalities due to COVID-19 between 2,647 counties with stay-at-home orders and 506 counties without orders on the same date. Although mean rates of change decline for all counties, the rightmost column shows that 1,931 of the 2,647 counties with orders are in groups that decline faster than counties without orders.

|  | Dependent Variable: One Week Change in Log Weekly Confirmed Cases | | | | | | Dependent Variable: One Week Change in Log Weekly Fatalities | |
|---|---|---|---|---|---|---|---|---|
|  | Model 1 7 Days After | | Model 2 14 Days After | | Model 3 21 Days After | | Model 4 21 Days After | |
|  | *Estimate* | *SE* | *Estimate* | *SE* | *Estimate* | *SE* | *Estimate* | *SE* |
| ***Difference Between Counties With and Without and Order, After the Order*** | -0.36 | 0.12 | -0.51 | 0.12 | -0.67 | 0.15 | -0.91 | 0.25 |
| ***Mean Difference Between Counties With and Without an Order*** | 0.38 | 0.09 | 0.38 | 0.12 | 0.34 | 0.14 | 0.64 | 0.26 |
| ***Mean Difference Before and After the Order*** | -0.24 | 0.07 | -0.27 | 0.15 | 0.19 | 0.13 | -2.07 | 0.00 |
| ***One Week Change in Log Weekly Tests Performed*** | 0.21 | 0.15 | 0.45 | 0.17 | 0.85 | 0.12 | -1.27 | 0.55 |
| ***Adjusted $R^2$*** | 0.82 | | 0.81 | | 0.85 | | 0.35 | |

**Table 3.** Difference-in-difference weighted least squares regression results. First row shows estimated treatment effect of stay-at-home orders on log weekly confirmed COVID-19 cases after 7 days (Model 1), 14 days (Model 2), and 21 days, and effect on log weekly COVID-19 fatalities after 21 days (Model 4). Results for all models are for two observations each (one on the date of the order and one after the order) for 2,647 counties grouped by date of order (*N* = 22), and paired with two observations each for 506 counties on the same dates where no order was issued. All 88 county-group observations are weighted by the number of counties per group, fixed effects for county group and date are included in each model and standard errors are clustered by group. These models control for all fixed factors that vary between counties and factors that vary over time at the national level.

spurious correlation between local factors (such as the date of onset of the disease or the capacity of the health system) and the timing of stay-at-home orders. To control for these factors, we apply a two-way fixed-effects difference-in-differences model to the data (see Methods).

Table 3 shows results from three models for growth in weekly confirmed cases that estimate the effect of stay-at-home orders 7 days, 14 days, and 21 days after the orders go into effect. A fourth model shows an estimate of the effect of orders on growth in weekly fatalities after 21 days. The key estimates are in the top row of Table 3 and exponentiating these coefficients $(exp(\beta) - 1)$ allows us to interpret them as percentage changes in weekly cases. They suggest that stay-at-home orders are associated with a 30.2 percent (11.0 to 45.2) reduction in weekly cases after one week, a 40.0 percent (23.4 to 53.0) reduction after two weeks, and a 48.6 percent (31.1 to 61.7) reduction after three weeks. Stay-at-home orders are also associated with a 59.8 percent (18.3 to 80.2) reduction in weekly fatalities after three weeks.

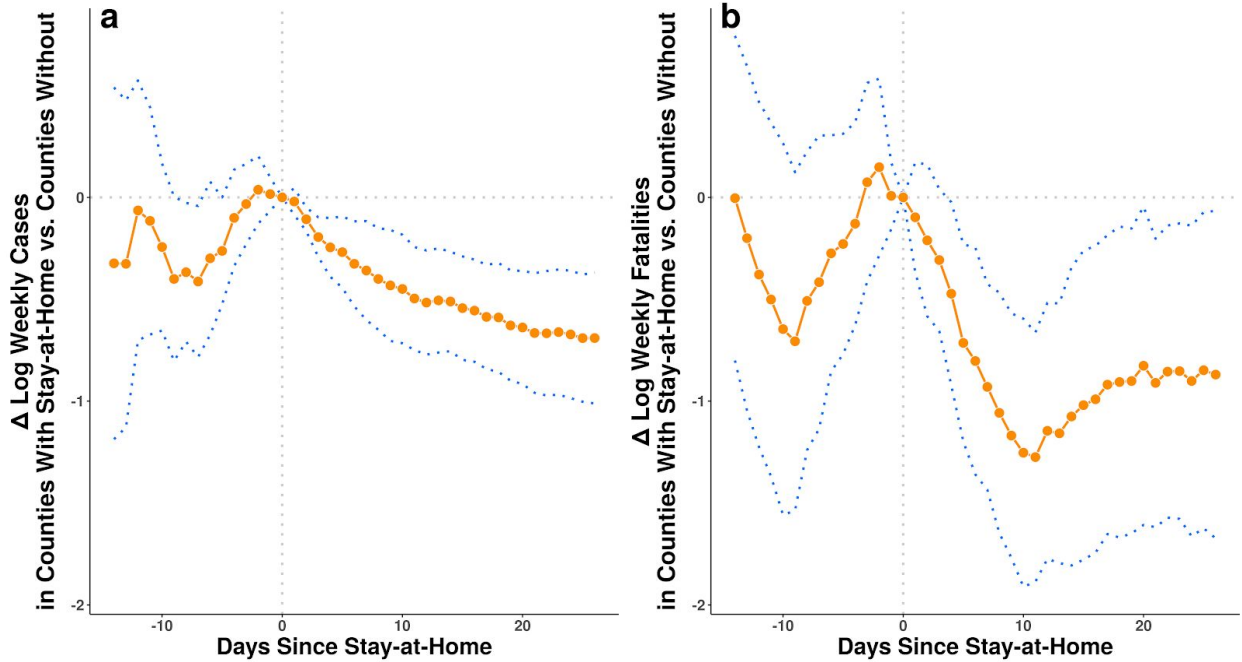

**Figure 4.** Estimated effect of stay-at-home orders on one week change in log weekly confirmed COVID-19 cases **(a)** and COVID-19 fatalities **(b)**, by the number of full days since the orders were issued. Estimates are derived from fixed-effects regression models that control for county-group level and date fixed effects and for correlated observations with cluster-robust standard errors at the county-group level (see Methods). Blue dotted lines are 95% confidence intervals.

Figure 4 shows these estimates along with the estimates for each day *prior* to and *after* the day a stay-at-home order goes into effect. The prior day estimates serve as a placebo test. Unlike the raw data shown in Tables 2 and 3, the estimates here are adjusted for unobserved factors that could influence the course of the disease that vary over time and between counties. Notice that the estimates in Figure 4 before the order goes into effect exhibit high variance and the confidence intervals span zero. This suggests that county-group-level differences in case and fatality growth are not strongly influencing the timing of stay-at-home orders, helping to rule out the possibility that the later associations we see are driven by reverse-causality or differential trends.

It is important to remember that one unit of a logarithm is an order of magnitude, and as such these results suggest stay-at-home orders had a large effect. We use the model estimates from Table 3 to produce a counterfactual number of unlogged cases and deaths for each county group under the assumption that growth in the absence of a stay-at-home order would have been higher by the estimated amount (see Methods). We employ the upper and lower confidence intervals of these estimates to calculate upper and lower bounds. These results suggest that stay-at-home orders reduced confirmed cases by 390,000 (170,000 to 680,000) and fatalities by 41,000 (27,000 to 59,000) within the first three weeks in localities where they were implemented.

## Discussion

The results here suggest that stay-at-home orders in the United States are effective in limiting the spread of COVID-19 and provide some hope that the physical distancing measures now widely implemented are working to "flatten the curve." The results also suggest that stay-at-home orders dramatically reduced both the number of cases and fatalities that result from the disease.

With that said, we note certain limitations in our analysis. Stay-at-home policies are ultimately assigned endogenously so, as with any observational study, we cannot say for certain that the associations we have measured are the result of a causal effect. Our tests of reverse causality suggest that stay-at-home orders influence case growth and not the other way around, but there is no way around the fact that these are observational data from which causal estimates are notoriously difficult to obtain.

Our dependent variables -- cases and fatalities -- are based on incomplete data. It is well-known that rates of testing in the United States were extremely low in the early part of the pandemic,[2,21] so measures of cumulative cases and fatalities over time probably increased faster than the disease itself due to the addition of previously undetected infections. And many cases and fatalities may have gone undetected entirely. We attempt to control for this issue with county and time fixed effects and a measure of the growth in testing, but we are unlikely to have entirely adjusted for local variation in access to tests.

Our independent variable, stay-at-home order status, measures a policy intervention that was often implemented simultaneously or within days of several other local interventions, such as bans on mass gatherings and closures of schools, non-essential businesses, and/or public areas. Given the uncertainty about how many days infected individuals are contagious both before and after the onset of symptoms, efforts to generate a sharp estimate of the effects of policies that were implemented within days of each other are difficult. Moreover, our analysis suggests these other local interventions might also have an effect on cases and fatalities. On average, the peak of infections happens three days *prior* to the stay-at-home order. In addition, we see significant reductions in the growth rate of cases and fatalities within days of the order. This is in spite of the fact that case identification during the early part of our observations was based on tests that often took a week to be resolved.[2,21]

With our current empirical approach we cannot perfectly separate the effects of other local interventions from that of stay-at-home orders. This means that our estimates should properly be interpreted as the effect of stay-at-home orders bundled with effects of correlated local interventions. Our controls are the set of counties that did not issue mandatory orders, but in those localities residents may have voluntarily changed their social behaviors. As such, our estimates are the difference between the effects of an average bundle of changes in counties that issued orders relative to an average bundle of changes in counties that did not. An interesting question which we leave for future work is which local interventions in the policy mix helped the most.

One final limitation is that we assume the effects of stay-at-home orders between localities are independent, but it is likely that significant spillovers exist. Consider the effect of the epidemic in New York City on neighboring counties in New Jersey, Connecticut, and as far away as Rhode Island. Or the effect of Mardi Gras in Louisiana or Spring Break in Florida on a variety of locales throughout the United States. This suggests significant spillover of infections. To the extent stay-at-home orders likewise spill over to other localities -- as seems reasonable to assume -- we have likely *underestimated* the effect of stay-at-home orders because control counties are partially treated by connections to counties with orders.

It is important to note that COVID-19 cases and fatalities continue to grow in many areas in the United States. Only when the rate of growth turns negative will we know whether or not we slowed the disease in time to keep it from overrunning the health system capacity in various localities. There is much still to be done, and we are hopeful that the work here will help our fellow scientists, policymakers, and the public-at-large to plan for the next steps in managing this disease.

## Methods

### Data

The time and date of county-level "stay-at-home" or "shelter-in-place" orders for each state and locality were aggregated and reported on a web page maintained by the *New York Times* starting on March 24, 2020.[23] As new orders went into effect, this page was updated. We checked it once daily to update the data through May 7, 2020. In some cases a statewide order was reported with reference to earlier city-level or county-level orders in the state without specifying where they occurred. In those cases, we searched local news outlets to find references to official city and county orders in the state that preceded the statewide order. For each county in each state we recorded the earliest time and date that a city, county, or statewide order came into effect.

County-level data on cumulative COVID-19 confirmed cases were also aggregated daily by the *New York Times*.[24] We discarded all observations where cases were not assigned to a specific county (these account for 0.8% of total cases). We retained observations where cumulative cases declined from one day to the next due to official revisions to the counts (0.8% of cases).

Availability of tests for COVID-19 in the United States has not been uniform over the date range of the study.[21] To mitigate the effect of changes in rates of testing on our measure of both confirmed cases and fatalities, we also collected data on the number of tests administered each day. This information is not currently available for each county, but it is available for each state by date from the COVID Tracking Project[25]. When we aggregated observations into county groups we measured the weighted mean of these state-level tests, where the weights are the number of counties from a county group in each state.

**Estimation**

In our data, we observe cases and fatalities $y$ for each county $k$ and date $t$. We would like to compare counties with orders to those without, so we aggregate cases and fatalities by county groups $c$ that each include all counties that implement the same stay-at-home policy. This includes 22 groups of counties that each chose to issue a stay-at-home order on a unique date and one group that never issued a stay at home order. Since both cases and fatalities grow exponentially in our data, we measure the rate of change as the difference in log weekly counts and we add 1 to ensure the log is defined:

$$\Delta y_{ct} = ln(\sum_{k \in c,\, t \in \{-6,0\}} y_{ct} + 1) - ln(\sum_{k \in c,\, t \in \{-13,-7\}} y_{ct} + 1) \ . \quad (1)$$

For each of the 22 county groups where an order was enacted ($x_c = 1$), we include in the data observations from two periods, one on the date of the stay-at-home order ($p_{ct} = 0$) and one on a date $d$ days in the future ($p_{ct+d} = 1$). For comparison, we also include two observations from the group that had no order ($x_c = 0$) on the same dates $t$ and $t + d$. This yields a total of 88 observations for each scalar value $d$ that we consider.

We model the effect of stay-at-home orders with a two-way fixed-effects weighted least squares regression model:[26]

$$\Delta y_{ct} = \alpha_c + \tau_t + \beta_1 x_c + \beta_2 p_{ct} + \beta_3 x_{ct} p_{ct} + u_{ct} \ . \quad (2)$$

A strength of this model is that fixed effects $\alpha_c$ control for all time-invariant features of each county group that might drive rates of change in cases and fatalities.[27] For example, each county has its own age profile, socioeconomic status, local health care system, base rate of population health, and date on which a first case of COVID-19 was observed. Additionally, time fixed effects $\tau_t$ control for factors that vary over time.[27] For example, case rates could be affected by changes in the availability of testing nationally, in social behaviors influenced by daily events reported in the media, and national-level policies that vary from one day to the next.

Including $\beta_1 x_c$ allows us to control for the overall difference between counties that ever had an order and those that did not. Similarly, $\beta_2 p_{ct}$ allows us to control for the overall difference between period 1 and period 0. Including the interaction of these two variables $x_{ct} p_{ct}$ allows us to estimate $\beta_3$, the difference in the differences between counties that had an order and those that did not. This estimate captures the causal effect of stay-at-home orders on cases and fatalities under one assumption. The assumption is that counties that implement orders on a specific date would have had similar changes in cases as counties that did implement orders if the implementing counties had not issued the order. This is the standard parallel trends assumption of difference-in-difference models.

Finally, we also weight each county group observation by the number of counties and we cluster standard errors $u_{ct}$ at the county group level. This adjusts the estimated standard errors for unobservable factors correlated within county groups and allows interpretation of the marginal effects at the county level.

We can examine the temporal dynamics of stay-at-home orders by repeating the regression model for different values of *d* days between the date of the order and the later post-treatment period. We can also let *d* be negative to see if there are systematic differences in cases and fatalities between counties with and without orders prior to the date orders go into effect. This allows us to test whether differences in cases and fatalities might cause changes in the date an order is enacted, rather than the other way around. Due to data availability constraints, the full range of possible days that we can model is $d \in \{-14, 26\}$. Estimates for these models for both cases and fatalities are shown in Figure 3.

We can also use the regression results to estimate a counterfactual number of cases and fatalities possibly prevented by stay-at-home orders. Since $\beta_3$ is the estimated difference in change in log weekly counts, the counterfactual difference in unlogged counts is simply $[exp(\beta_3) - 1]$ times the observed weekly count. We calculate this separately for each county group for week 1 using the model where *d* = 7 and we repeat for week 2 (model *d* = 14) and week 3 (model *d* = 21), summing over all 22 county groups and all 3 weeks.

**Contributors**

All authors contributed to collection of data, design and execution of analysis, and drafting, review, revision, and approval of the final manuscript.

**Competing interests**

The authors declare no competing interests.

**Acknowledgments**

We thank John Ahlquist, Robert Bond, Lawrence Broz, Manuel Cebrian, Chistopher Dawes, Scott Desposato, Chris Fariss, Anthony Fowler, Jeffry Frieden, Richard Garfein, Micah Gell-Redman, Gary Jacobson, Lauren Gilbert, Tim Johnson, Arman Khachiyan, Thad Kousser, Richard Kronick, Sam Krumholz, Brad LeVeck, Natasha Martin, Peter Loewen, Lucas de Abreu Maia, David Meyer, Robyn Migliorini, Stan Oklobdzija, Niccolò Pescetelli, Daniel Rubenson, Wayne Sandholtz, Chip Schooley, Zachary Steinert-Threlkeld, Tracy Strong, Clara Suong, Davide Viviano, and Harla Yesner for helpful comments.